# Stand for Something or Fall for Everything: Predict Misinformation Spread with Stance-Aware Graph Neural Networks

**Zihan Chen**
Stevens Institute of Technology
Hoboken, NJ, USA
zchen61@stevens.edu

**Jingyi Sun**
Stevens Institute of Technology
Hoboken, NJ, USA
jsun54@stevens.edu

**Rong Liu**
Stevens Institute of Technology
Hoboken, NJ, USA
rliu20@stevens.edu

**Feng Mai**
Stevens Institute of Technology
Hoboken, NJ, USA
fmai@stevens.edu

## Abstract

*Although pervasive spread of misinformation on social media platforms has become a pressing challenge, existing platform interventions have shown limited success in curbing its dissemination. In this study, we propose a stance-aware graph neural network (stance-aware GNN) that leverages users' stances to proactively predict misinformation spread. As different user stances can form unique echo chambers, we customize four information passing paths in stance-aware GNN, while the trainable attention weights provide explainability by highlighting each structure's importance. Evaluated on a real-world dataset, stance-aware GNN outperforms benchmarks by 32.65% and exceeds advanced GNNs without user stance by over 4.69%. Furthermore, the attention weights indicate that users' opposition stances have a higher impact on their neighbors' behaviors than supportive ones, which function as social correction to halt misinformation propagation. Overall, our study provides an effective predictive model for platforms to combat misinformation, and highlights the impact of user stances in the misinformation propagation.*

## Introduction

The rapid growth and widespread adoption of social media platforms have revolutionized the way people communicate, access information, and form opinions. While these platforms offer numerous benefits, they also facilitate the spread of misinformation that can have far-reaching consequences on public opinion, decision-making, and social behavior (Friggeri et al., 2014; Oh et al., 2013). The prevalence of misinformation on social media platforms has emerged as a significant challenge for researchers, policymakers, and platform administrators (Lazer et al., 2018; Vosoughi et al., 2018).

Proactively identifying users predisposed to propagating misinformation is vital in mitigating its negative effects (Kim et al., 2019; O'Connor & Weatherall, 2019). Traditional platform interventions primarily rely on content moderation, restriction policies, and fact-checking tools from third-party websites. For example, Twitter has applied warning labels on misleading content and promoted authoritative information, while Facebook has collaborated with independent fact-checking organizations to review and rate the content accuracy on its platform. Nevertheless, these measures have demonstrated limited effectiveness in curtailing misinformation spread. Existing studies show that misinformation flags can heighten users' cognitive awareness, induce cognitive dissonance, and emphasize their limited ability to discern misinformation (Moravec et al., 2019, 2022). Moreover, platform-driven misinformation tagging can inadvertently impact untagged content. For instance, Pennycook et al., (2020) find that when fewer than one-third of false headlines are tagged, the likelihood of users spreading misinformation may unintentionally increase. These limitations underscore the need for novel approaches to address the pressing challenge of misinformation dissemination.

In this study, we leverage the users' stances and their influence on others to predict the misinformation spread. Stance represents a user's articulated position or opinion towards a specific target, such as an individual, organization, or issue. It is crucial to differentiate stance from "sentiment," which centers on identifying the overall emotional tone of a text rather than pinpointing the user's viewpoint concerning a subject. Although the two terms are frequently—and inaccurately—used interchangeably, understanding their distinct roles is vital for the online community analysis. A majority of prior information systems (IS) studies focus on the influence of misinformation on users' emotions and the subsequent effects on misinformation dissemination (Deng & Chau, 2021; Horner et al., 2021; Kim & Dennis, 2019). However, limited attention has been given to the impact of users' stances on misinformation propagation. In our user-post scenario, consider the following posts as an illustration:

- Original post: As deep learning continues to gain popularity, Google CEO warns society to brace for AI acceleration.

- User 1's post: Honestly, the thought of AI acceleration is terrifying, but we can't ignore the benefits. We must embrace and support this development to harness its full potential.

- User 2's post: The advancements in AI and deep learning are truly amazing! But I strongly believe we need to halt AI acceleration for now. We must prioritize safety and ethics before speeding ahead.

This example highlights the important differences: User 1's post exhibits negative sentiment but supports the original post, while User 2's post shows positive sentiment but opposes the original post. Thus, compared with sentiment, stance provides a more granular understanding of users' positions within an interaction network, leading to better insights into online community dynamics and opinion evolution.

Furthermore, we employ user stances to study the polarization and echo chambers existence, which tracks the formation of clusters that share similar opinions. This focus aligns with an emerging stream of research that highlights the importance of users' social connections and their neighbors' attitudes in shaping their behaviors (Moravec et al., 2020). However, the role of user stance in forming and modifying echo chambers—a prominent social dependency mechanism that fuels misinformation spread (McPherson et al., 2001)—remains relatively unexplored. Consider a user who has multiple followers: if the user supports a piece of misinformation, his/her followers are exposed to this affirmative stance, which increases the likelihood that they adopt a similar position and thereby facilitate the spread of false information. Conversely, if the user expresses a strong opposition to the misinformation, this could serve as a form of "social correction" (Vraga & Bode, 2017), making his/her followers more likely to inhibit its dissemination. Overall, a polarized user's stance can act as a double-edged sword: it can either exacerbate or mitigate the spread of misinformation, depending on the orientation of their views.

To address this, we propose a Stance-Aware Graph Neural Network (stance-aware GNN) that treats user information and post content as node features, and interactions as edges. As a state-of-the-art deep learning model proficient in handling graph structures, GNN embeds nodes into low-dimension vectors that retrains both nodal content and network topology (Wu et al., 2021). In the case of predicting misinformation spread, GNN embeds three key factors: users' characteristics, misinformation content, and network structure that amplifies users' attitudes and influence (Turel & Osatuyi, 2021). Additionally, as different user stances give rise to distinct echo chamber structures, they create diverse information/message passing paths for GNN, thus leading to varying embeddings. Stance-aware GNN subsequently aggregates these embeddings using trainable attention weights. By examining the attention weights after model training, we can identify which structures play a more critical role in the misinformation spread predictions.

To evaluate the performance of our model, we conduct experiments using a real-world dataset comprising 21 million tweets from 26,000 threads, linked to 13,000 fact-checked claims (Nielsen & McConville, 2022).

We select benchmark models that utilize text features, network features, or a combination of both to train classifiers. The comparison of their predictive capabilities is based on the area under the receiver operating characteristic curve (AUC) on the test data. Experimental results demonstrate that our stance-aware GNN significantly outperforms traditional benchmark models, achieving a 32.65% improvement in AUC. Moreover, our model surpasses GNNs without user stance and social correction by at least 5.01%. These findings highlight the effectiveness of user stances and their associated network structures in augmenting the prediction framework of misinformation spread.

The primary contributions of our study are threefold. First, we examine the influence of user stance on the formation of echo chamber structures and the subsequent misinformation diffusion. By proposing four types of information propagation paths based on users' relationships (i.e., following, co-sharing) and their polarized stances (i.e., support, opposition), our research highlights the importance of ordinary users' attitudes in countering misinformation, particularly in cases where platform interventions are of limited effectiveness.

Second, we introduce stance-aware GNN as a theory-driven deep learning model to examine the spread of misinformation. In contrast to the "black-box" models with limited transparency (Benbya et al., 2020), the four paths we proposed not only define a schema for GNN to aggregate information, but also serve as a lens to scrutinize our hypothesized mechanisms underlying how user-level behaviors can either accelerate or curb the misinformation dissemination. The attention weights of stance-aware GNN further provides explainability by showing that opposing user stances are more crucial than supporting stances for accurate model predictions. These findings align with recent studies suggesting that ordinary users' attitudes can act as social correction to curtail the dissemination of misinformation (Borwankar et al., 2022). Furthermore, from a model design standpoint, our model is the first model that proposes to use stance as the edge feature of GNN, and also the first that pinpoints the important differences between stance and sentiment when integrating them with GNN.

Third, our model carries significant practical implications. Evaluated on a real-world Twitter dataset, stance-aware GNN significantly outperforms all benchmarks with improvements ranging from 81.99% to 4.69%. The performance demonstrates the potential of stance-aware GNN as a valuable tool for platforms and organizations aiming to proactively mitigate the misinformation effects, even when faced with the challenges of processing substantial data volumes on modern platforms. This impressive efficacy also underscores its aptitude in identifying the specific user subset most susceptible to misinformation. Its direct application would be empowering platforms to design targeted interventions, thus diminishing the probability of users being swayed by false narratives. It also provides stakeholders with a robust tool to assess the potential repercussions of misinformation and gauge its dissemination scope.

The remainder of this paper is organized as follows. In the next section, we offer an overview of related work on misinformation spread mitigation, echo chamber mechanisms in networks, and deep learning techniques in information systems. We then delve into the details of our proposed stance-aware GNN, including the identification of user stance, construction of network structures for message passing, and development of GNN for prediction. Next, we present the experimental setup, results, robustness test, and discuss the model explainability. We close by concluding the paper, pinpointing the potential limitations, and proposing avenues for future research.

## Theory and Related Literature

### Misinformation Dissemination Mitigation

Existing research in Information Systems (IS) has examined the effectiveness of platform interventions such as fake news flags (Moravec et al., 2019; Ng et al., 2021; Ross et al., 2018), source rating (Kim et al., 2019; Kim & Dennis, 2019; Moravec et al., 2022), and forwarding restriction policies (Ng et al., 2021). The findings from these studies show that platform interventions can increase users' cognition levels but may have limited effectiveness in combating misinformation spread. For example, studies show that fake flags can create cognitive dissonance in users that triggers them to realize their automatic cognition in social media information consumption and instead invest more cognitive effort in processing information (Moravec et al., 2019). However, with increased levels of cognition, users are unlikely to change beliefs or be more capable in distinguishing fake news from truth (Ross et al., 2018). Furthermore, studies show that source ratings, which indicate the reputation of information sources, can remind users to think critically about the information (Kim & Dennis, 2019). Specifically, asking users to rate stories can effectively improve users' cognitive levels because it reminds users that they may lack the knowledge to distinguish trustworthy sources (Moravec et al., 2022). Although source ratings can affect users' perception of the believability of information, it is unclear how that can affect actual sharing behaviors. In addition, flagging may have unintended consequences such that it affects the way users comprehend unflagged content. Once users have seen news headlines with fake flags, they tend to see unflagged headlines as more accurate (Pennycook et al., 2020). Experiments show that if below one-third of false headlines are tagged, the net effect of implementing warnings may even increase the spread of misinformation (Pennycook et al., 2020).

Given the challenges in top-down platform interventions, an emerging stream of studies start to focus on the role of ordinary users. On social media platforms, users can also make corrections when commenting on posts containing misinformation, and such correction can be known as "social correction" (Vraga & Bode, 2017). In experimental studies, social correction was found to be equally effective in reducing misperceptions compared to platform corrections through authoritative sources (Vraga & Bode, 2017). This is somewhat challenged in another study by Kim et al. (2019), who compared three types of ratings: expert ratings of articles, user ratings of articles, and user ratings of sources, and their results show that expert ratings and user ratings of articles had stronger effects on believability than users source ratings. This finding suggests that ordinary users' ratings of news articles can also be effective, but they generally lack the knowledge to evaluate sources as effectively as experts (Kim et al., 2019). To extend studies in laboratory settings, recent studies leverage the Twitter Birdwatch program to gauge the effect of crowdsourced misinformation monitoring (Borwankar et al., 2022; Saeed et al., 2022). In particular, this program randomly selected ordinary users to write misinformation notes for posts and such notes were visible to other users. Studies show that such crowd-based misinformation intervention resulted in decreased spread of misinformation and sentiment extremity (Borwankar et al., 2022). In other words, "wisdom of crowd" can be leveraged from ordinary users to remind others about problematic information. In sum, these recent studies suggest that ordinary users' opinions may be a scalable solution to misinformation dissemination.

### Echo Chamber in Misinformation Dissemination Networks

The role of social correction in misinformation dissemination networks warrants investigation due to its integral connection with information diffusion processes, which are primarily influenced by social dependencies. A key mechanism driving misinformation propagation is the echo chamber mechanism, where individuals tend to seek and engage with information that reinforces their pre-existing beliefs (Shore et al., 2018). Echo chambers are formed by homophily (McPherson et al., 2001) and confirmation bias (Nickerson, 1998), fostering individuals to be exposed to like-minded others (Bakshy et al., 2012). Echo chambers foster the formation of fragmented groups that resonate rather than challenge one another, thereby limiting the exposure to diverse opinions (Kitchens et al., 2020; Shore et al., 2018).

In social network terms, echo chambers can be represented as "echo" (homophily) and "chamber" (network closure) (Jasny et al., 2018; Shore et al., 2018). Specifically, the "echo" mechanism denotes information exchange between actors sharing the same views on an issue (Jasny et al., 2018) or disseminating information akin to that received from followers (Shore et al., 2018). The "chamber" mechanism represents a closed social network neighborhood where ideas reverberate within a small, tightly-knit group isolated from external influences (Shore et al., 2018). The structural tendency towards homophily and closure can be explained by balance theory (Heider, 1982), which posits that actors are more likely to form close communities like-minded individuals to alleviate cognitive dissonance (Festinger, 1962). Recent studies on misinformation spread show that echo chambers account for the sharing of highly polarized political information (Shore et al., 2018; Song et al., 2020) and conspiracy theories (Xu et al., 2021).

Considering the impact of ordinary users' opinions and their influence on others in misinformation diffusion, it may be problematic to ignore their stances in the formation of echo chambers. To distinguish the misinformation consumers based on stances, they can be categorized into propagators, endorsers, and resistors (George et al., 2021). Propagators simply share information without commenting but such sharing is an indication of implicit agreement. Endorsers express explicit agreement with the misinformation shared through making confirmative comments. Resistors are fact-checkers who identify and oppose misinformation through active refutation against misinformation. These different roles are reflected as stances in user-post interactions, and such stances are likely to influence the information sharing behaviors of other users who are socially close. Consequently, user stance can alter the structure of echo chambers.

Following prior studies, we define the "echo" mechanism as users sharing the same information they receive from individuals they follow (Shore et al., 2018). Incorporating user stance, we suggest that echo chambers are more likely to form when users adopt a supportive stance, and less likely when they adopt an objective stance. That is, if User 1 shares misinformation with a supportive stance, User 2, who follows User 1, is more inclined to share the same information, given their pre-existing connection implies a certain degree of shared perspectives. Conversely, if User 1 debunks misinformation while sharing, homophily may be disrupted because User 1's negative stance will cause cognitive dissonance in User 2 (Festinger, 1962). To restore balance, User 2 can either reduce the identification with User 1 as an ingroup member (e.g., unfollowing User 1), or change opinions and refrain from sharing that information. The rationale is that users who lack support from their ingroup members are likely to experience a strong need to reduce cognitive dissonance through attitude change (McKimmie et al., 2003). Thus, User 2 is less likely to share the misinformation post if the followee User 1 debunks it.

Likewise, the "chamber" mechanism refers to users' clustering through repeatedly sharing the same set of information, forming closed and segregated groups. Neighborhood clustering is represented by closed triplets in one-mode networks and four-cycles in two-mode networks (Monge & Contractor, 2003). In our context of heterogeneous graphs involving both information and users, four-cycle refers to repeated co-sharing of the same set of information. These shared mindsets, exhibited through prior engagement with the same information, increase the likelihood of co-sharing new information in the future. Misinformation can diffuse within such network clusters because social support from confederates provides cognitive consonance (Stroebe & Diehl, 1981), which reduces the need for attitude change upon encountering fact-checking information. However, if the source of social support (i.e., User 1) challenges misinformation while sharing, User 2 may experience increased cognitive dissonance and decreased social support. To restore cognitive balance, User 2 can either distance themselves from the clustered group or change their attitude to maintain consonance with User 1. Either way, User 2 is less likely to interact with that piece of information. In summary, echo chamber mechanisms can exhibit different structures when incorporating stances into the network mechanisms through which misinformation spreads.

### *Deep Learning and Graph Neural Network in IS*

Our proposed approach is related to a burgeoning body of IS research that leverages machine learning (ML) techniques to mitigate its adverse effects. For instance, King et al. (2021) investigate the propagation dynamics of misinformation and corrective messages on social media platforms. Utilizing panel vector autoregression and a Twitter dataset, the authors reveal a bidirectional interplay between the false and true information. In a similar vein, Clarke et al. (2021) employ six distinct ML algorithms to identify financial related fake news. By analyzing linguistic features extracted from news articles, their study demonstrates the effectiveness of ML models in neutralizing deceptive narratives aimed at manipulating market conditions. In recent years, deep learning (DL) models have gained prominence as powerful computational tools capable of processing substantial data volumes on online platforms (LeCun et al., 2015). Owing to their ability to learn data representations across multiple levels of abstraction, DL models alleviate the need for time-consuming feature engineering, thereby enhancing the efficiency of tackling the misinformation dissemination (MA et al., 2016).

Despite the promising performance exhibited by DL models in analyzing misinformation propagation, two significant gaps persist in the current body of literature. First, misinformation does not propagate in isolation but within dynamic networks, where individuals' behaviors are strongly influenced by others' attitudes (Moravec et al. 2020). Furthermore, the diffusion of multiple pieces of misinformation can interact, either inhibiting or amplifying one another, which presents additional challenges for existing algorithms (Yoo et al., 2019). To address this issue, we employ Graph Neural Network (GNN), a state-of-the-art DL technique adept at handling the complexities of network-based tasks (Wu et al., 2021). GNNs extract high-level representations of each node based on its neighboring characteristics and network structures, yielding node embeddings that retain both network topology and content information (Hamilton et al., 2017; Kipf & Welling, 2017). Our study represents the social network as a heterogeneous graph with users and posts as nodes, where GNN learns their embeddings in an end-to-end manner based on the subsequent prediction task.

Second, while most existing DL methods are effective at detecting misinformation, they often lack transparency in explaining how models reach their outputs (Benbya et al., 2020). In accordance with the above theoretical development, we define four paths to represent "echo" and "chamber" mechanisms with supportive and opposed stances. Specifically, in the homophily mechanism where users share the same information as they receive from other users they follow, we compare two paths where the followee user holds positive and negative stances. Similarly, in the clustering mechanism where users repeatedly co-share the same set of information, we compare two additional paths where one user holds positive and negative stances. If user stance can influence the formation of network mechanisms through which misinformation spreads, capturing these stance-based echo chamber configurations should significantly improve GNN's performance when predicting misinformation spread. More importantly, we incorporate attention weights to emphasize the relative importance of each path, elucidating the mechanisms underlying the embedding process and enhancing the model's accountability.

In the realm of IS research utilizing GNNs, our work is related to Ma et al. (2021) that employs GNNs and metapath to predict user interactions with brand posts on social media. While their study offers valuable insights into the GNNs' effectiveness on heterogeneous graphs, it overlooks users' influence as well as their attitudes towards the posts. In light of Turel and Osatuyi's (2021) assertion that social media peers are highly efficient at influencing individuals' beliefs and behaviors, our research delves deeper into how users' attitudes can impact others and generate corresponding information propagation paths. Another pertinent study is Yuan et al. (2021) which employs GNNs to assess the authenticity of news articles by analyzing news content, images, and domain knowledge as features. Although their approach provides a solid foundation, it considers all nodes as homogeneous entities. Our work expands upon this by incorporating various node types and accounting for diverse interactions between them. This enhancement not only more accurately reflects real-world scenarios but also introduces additional features that bolster model performance.

# Method

The stance-aware graph neural network (stance-GNN) comprises two main components: the echo chamber structures based on users' stances, and their incorporation into an explainable GNN model. In this section, we first outline the method for identifying users' stances while they interact with messages. Then we delve into the stance-GNN structure, the integration of information passing paths, and the model's capacity for explaining paths' functionalities.

## *Labeling Stances with the Fine-tuned Model*

Though numerous IS studies have explored the impact of misinformation on users' sentiments, a gap remains in examining the significance of users' stances in this context. As outlined in the introduction, stance and sentiment play differentiated roles in online interactions. For instance, a user might express negative sentiment yet endorse the misinformation, or conversely, manifest positive sentiment but oppose the given post. To bridge this gap, we need a robust stance labeling system. We start the process by collecting labeled stance datasets to train and evaluate a stance detection model.

Stance data can be categorized into two types. The first type consists of sentence pairs and each with an associated stance label, originating from datasets such as FNC-1, Semeval2017, Argument Reasoning Corpus, and RumourEval2019 (Gorrell et al., 2018; Habernal et al., 2018; Rosenthal et al., 2019). The second type encompasses sentences directed towards specific topics (e.g., Biden/Trump), including SemEval2016 and WTWT (Conforti et al., 2020; Mohammad et al., 2016). Given that online posts typically consist of sentences, we adapt topic-based sentences by rephrasing them to align with our objectives. For instance, we convert a topic "Trump" from the presidential election dataset into a sentence: "The topic of this sentence is about Trump during the presidential election." We tailor the rephrasing according to each dataset's characteristics, ensuring the inclusion of all pertinent information. We adhere to the original papers' guidelines for train-test splits, and retain only support, against, and neutral stance labels. Unrelated labels are omitted, as user-tweet interactions typically possess some relevance.

In terms of the modeling approach, we select RoBERTa, a robust deep learning model renowned for its strong transfer learning capabilities in NLP tasks, over training a model from scratch (Liu et al., 2019). We employ the RoBERTa version pretrained on the MultiNLI dataset to capitalize on the embedded sentence information. Subsequently, we fine-tune the model with aforementioned training data. The final stance labeling model achieves an accuracy of 82.3% on the test set, representing a significant improvement over the 60.1% accuracy of RoBERTa without fine-tuning. Owing to its promising performance on sentence pairs, we employ this fine-tuned model to label stances in our empirical experiment.

## *Constructing Information Passing Paths*

As highlighted in the literature review, different stances from users can form distinct structures of echo chambers, subsequently impacting the misinformation diffusion (George et al., 2021). To account for these user stances (i.e., supportive, oppositional, or neutral) and the resulting information propagation patterns, we introduce two categories of network structures as information passing paths: Follower-based Network Structure (i.e., homophily) and Engagement-based Structure (i.e., clustering). These two categories, together with supportive and opposing stances, form four distinct information propagation paths within the network, as depicted in Figure 1.

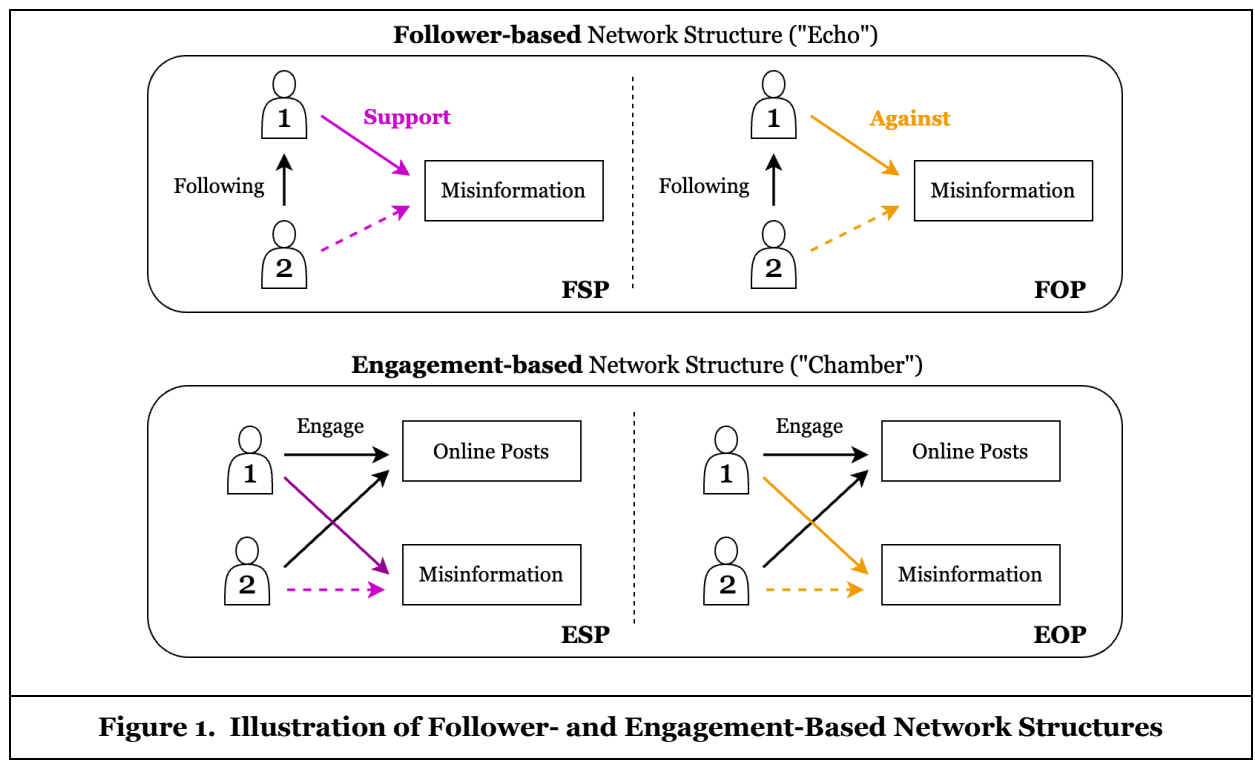

**Figure 1.  Illustration of Follower- and Engagement-Based Network Structures**

**Follower-based Network Structure**: The follower-based network structure can be classified as two information propagation paths: Follower-based Support Path (FSP) and Follower-based Opposition Path (FOP).

- Follower-based Support Path (FSP):  In this path, User 1 is followed by User 2. Further, User 1 supports the misinformation when encountering an online misinformation post. Consequently, it is likely that User 2 will support the misinformation and contribute to its spread.
- Follower-based Opposition Path (FOP): Similarly, User 1 is followed by User 2, but User 1 opposes the online misinformation post. As a result, it is likely that User 2 will oppose the misinformation and refrain from spreading it.

Follower-based paths represent direct connections between users. Previous research demonstrates that users connected through following relationships tend to exhibit homophily, with individuals sharing similar beliefs, preferences, or opinions following each other (McPherson et al. 2001). Likewise, users' prior engagement with the same content also reflects shared interests and forming communities, which serves as an indirect, weaker tie among users. Research further indicates the importance of weak ties for disseminating information to broader audiences (Bakshy et al. 2012). Based on these findings, we propose a second type of structure, Engagement-based network structure, and its associated propagation paths.

**Engagement-based Network Structure**: The engagement-based structure includes two paths for information propagation: Engagement-based Support Path (ESP) and Engagement-based Opposition Path (EOP).

- Engagement-based Support Path (ESP): In this path, User 1 and User 2 have previously engaged with the same post. When encountering a new misinformation post, User 1 supports the misinformation. Consequently, User 2 is more likely to support the misinformation and contribute to its dissemination.
- Engagement-based Opposition Path (EOP): User 1 and User 2 have previously engaged with the same post. If User 1 opposes a new misinformation post, we posit that User 2 is also more likely to oppose the misinformation and abstain from spreading it.

We elaborate further on the construction of the four path types, using data from social media platforms as the example context. First, we identify the focal user who interacts with misinformation and determine the stance of this interaction. For the follower-based structure, we identify the focal user's followers, positing that they would be exposed to the user's behavior via following. If this focal user supports (opposes) the misinformation, their endorsement (denouncement) should amplify (reduce) the misinformation's reach. Consequently, we establish an edge between the followers and the misinformation, giving rise to the FSP (FOP) structure. For the engagement-based structure, we identify users who previously engaged with other posts (e.g., replying, commenting, liking, or sharing) alongside with the focal user. If the focal user supports (opposes) the misinformation, this endorsement (denouncement) can further amplify (reduce) its dissemination through weak ties based on past co-engagements. In response, we create a linkage between those who have co-engaged with the focal user and the misinformation, leading to the formation of the ESP (EOP) structure.

The four information propagation types account for both direct social network ties, represented by user following, and indirect ties, represented by shared engagement with the same post. By examining these ties, we can gain insights into how information flows through networks and how individuals are influenced by their socially proximate others. These predefined structures also enable us to embed well-established echo-chamber theories into deep learning models, thereby facilitating a theory-driven approach to model design. In the following section, we describe the integration of these four path types into our GNN model.

### Stance-Aware Graph Neural Networks

In this section, we present a GNN model that incorporates the four aforementioned information passing paths. We start by constructing a heterogeneous graph to represent user-post interaction networks. A heterogeneous network can be defined as $G = (V, E)$, where $V$ and $E$ represent nodes and edges, respectively. There are two node types in our case: users, denoted by $U$, and online posts, denoted by $P$. The features of a user, such as self-description, number of posts, and account age, can be represented as $h_{u_i}$. Similarly, the features of a post, such as its text content, can be represented as $h_{p_i}$. Concerning the edge, we have three edge types: edges between users (e.g., following, mentioning), edges between posts (e.g., shared keywords), and edges between users and posts (e.g., posting, quoting).

Our objective is to predict whether a user will spread misinformation in the subsequent time period. This implies that our model can only utilize the graph and features at time t for training and then predict user-misinformation interactions in the next time period. Consequently, we denote the training time period as t and the prediction horizon as $L$. The graph structure that evolves over time is represented as $G_t$ and $G_{t+L}$. Our target is to predict user-misinformation interactions at time $t + L$. Following this, we employ the GNN to embed each node into a vector space, which integrates both network-level features and node-level features from the node itself and its neighbors. To simplify, we describe the embedding of GNN as a three-steps process: message passing, weighted aggregation, and feature updating. We will use User 1 and associated features as an example to explain the embedding process.

During the message passing phase, the GNN acquires features from User 1 and his/her neighbors from the previous layer (denoted as layer $l$). If the layer is the first one (i.e., $l = 1$), then the GNN will directly pass the original nodal features. Subsequently, these features are multiplied by a trainable weight matrix and passed through an activation layer (e.g., ReLU), which means that the messages from User 1's neighbors are weighted and aggregated together. Finally, the GNN updates User 1's features with the aggregated result. This process can be expressed using the equations below.

$$m_{N(u_1)}^{(l)} = AGGREGATE^{(l)} \left( \left\{ h_v^{(l)}, \forall v \in N(u_1); \Theta_{AGG}^l \right\} \right) \tag{1}$$

$$h_{u_1}^{(l+1)} = UPDATE^{(l)} \left( h_{u_1}^{(l)}; m_{N(u_1)}^{(l)}; \Theta_{UPD}^l \right) \tag{2}$$

where $\Theta_{AGG}^l$ and $\Theta_{UPD}^l$ represent the trainable parameters of the aggregation function and update function at layer $l$ respectively. $v$ denotes the neighbor of User 1, and $m_{N(u_1)}^{(l)}$ is the "message" that is aggregated from $u_1$'s graph neighborhood $N(u_1)$. As a result, the embedding of User 1 from the GNN is denoted as $h_{u_1}^{main}$, where $main$ indicates that the embedding originates from the entire network. Similarly, the GNN embedding of each post, such as Post 1, is denoted as $h_{p_1}^{main}$.

As evident from the above explanation, the output of the GNN is highly correlated with the selection of neighbors. In other words, if the neighbors are different, the embeddings will also differ significantly. Thus, to integrate the four types of echo chamber structures into the stance-GNN, we select specific neighbors associated with each information propagation path, which consequently produces varying embeddings even for the same node. Specifically, we first construct the network based on the neutral stance between nodes as message passing paths, which retain the basic network structure. Then, for every network structure type, we insert the type into the network based on the existence of the structure. For instance, in the case of FOP, if User 2 follows User 1 and User 1 displays an opposing stance towards misinformation, an opposing path will be established between User 2 and misinformation to represent the potential opposing attitude of User 2.

Continuing with User 1 as an example, incorporating the four types of paths will yield four neighbor sets and consequently, four distinct embedding results. We denote the embeddings of User 1 from the main GNN and four path types as $H'_{u_1} = [h_{u_1}^{main}, h_{u_1}^{fsp}, h_{u_1}^{fop}, h_{u_1}^{esp}, h_{u_1}^{eop}]$. However, as previous research indicates, not all propagation paths are useful, and some may introduce noise that could negatively affect model performance.

To address this issue, we design a trainable attention weight mechanism to aggregate the embeddings from the four paths. As the attention weights are trainable, the model can identify which weights contribute significantly to its prediction, and adjust their values accordingly. In essence, a high attention value indicates the importance of a feature, while a low value implies its insignificance. This provides explainability for our model, allowing us to determine the effectiveness of each message passing path by examining the final attention scores. The aggregation process can be represented as:

$$h'_{u_1} = w_{Att} H'_{u_1} \qquad\qquad h'_{p_1} = w_{Att} H'_{p_1} \tag{3}$$

where $w_{Att} \in R^{1 \times 5}$ is the aforementioned trainable attention. The aggregation of post embeddings uses the same attention weight, as the attention scores only reflect the relative importance among paths and are thus shared between user and post.

Next, as the spread of misinformation involves both user and post, we concatenate the user and post embeddings and input them into a final classifier. Here, we use a multilayer perceptron model as a simple classifier. The output is then passed through a sigmoid function to represent the probability that the user will spread the misinformation (i.e., $\hat{y} = 1$). The output layer can be represented as:

$$h'_{u_1-p_1} = [h'_{u_1}, h'_{p_1}] \qquad (4) \qquad \qquad \hat{y}_{u_1-p_1} = Sigmoid\left(MLP(h'_{u_1-p_1}; \Theta_{MLP})\right) \qquad (5)$$

where $\Theta_{MLP}$ represents the trainable parameters in the output layer, $MLP$ denotes fully-connected neural networks, and $MLP$ stands for the sigmoid function. The comparison of prediction (i.e., $\hat{y}$) and ground truth (i.e., $y$) generates the loss. The loss is then backpropagated through the model to adjust the parameters, optimizing the model's performance.

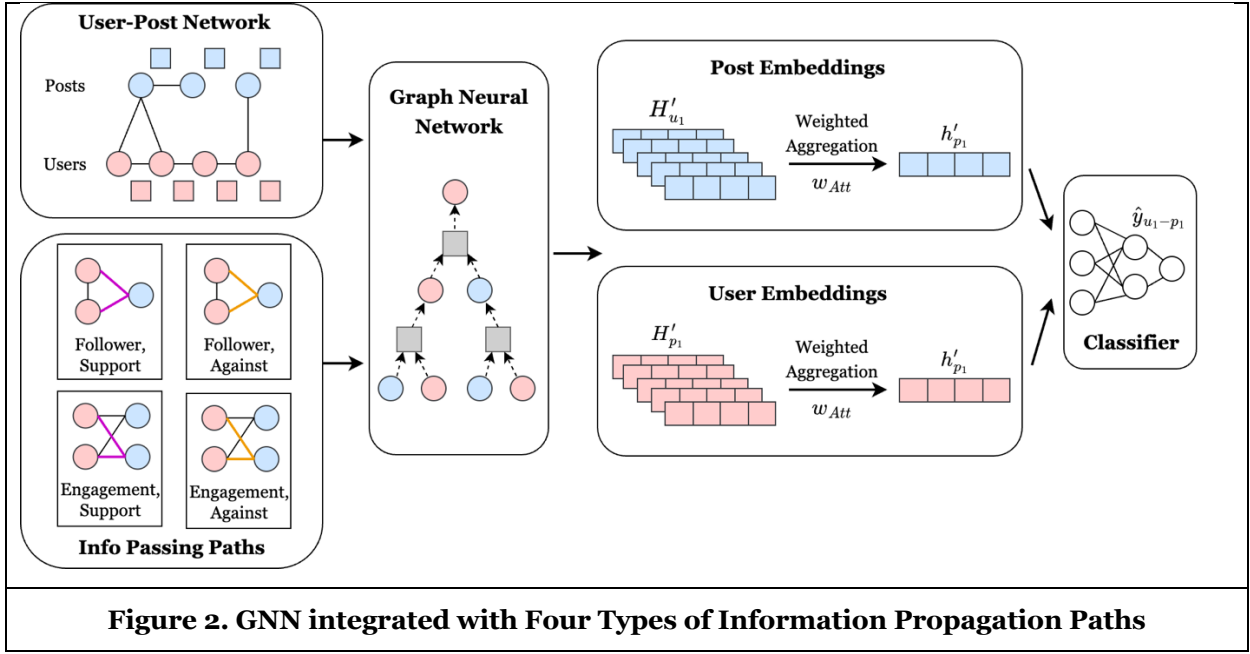

**Figure 2. GNN integrated with Four Types of Information Propagation Paths**

Our objective is to predict whether a user will propagate misinformation in a subsequent time period. On social media platforms like Twitter, propagate behaviors can include a user posting, retweeting, quoting, replying to, or mentioning the misinformation post, which promotes misinformation spread through their affirmation (George et al., 2021). If such a link between the user and the misinformation emerges in the ensuing period, we label it as 1, otherwise 0. This means that our model performs a binary classification task. Accordingly, we employ the Binary Cross Entropy as the loss function to compare prediction (i.e., $\hat{y}$) with ground truth (i.e., y) and generate the loss. The loss is then backpropagated through the model to adjust trainable parameters, optimizing the model's performance.

In summary, the model above captures the key factors essential for predicting misinformation spreading behavior: the node features representing user characteristics and post content, and the message passing representing the network structure. Moreover, the four types of information passing paths reflect the echo chamber structures determined by users' stances, reflecting how the ordinary users can sway their neighbors' attitudes and thus affect misinformation dissemination. These stance-based paths along with the GNN embed transfer the original features of users and posts ($h_{u_i}$, $h_p$.) to the embedded features ($h'_{u_i}, h'_{p_i}$), which represent aggregated features that encapsulate both nodal and topological information. Finally, the attention weights indicate the functionality and effectiveness of each path, providing explainability for the stance-aware GNN model.

## Experiment

We commence our experiment by utilizing the MuMiN dataset to evaluate our model's performance (Nielsen & McConville, 2022). The original dataset consists of three subsets with varying sizes, totaling 13,000 fact-checked claims and 21 million tweets across 26,000 Twitter threads. We initially focus on the largest version, as it offers the most comprehensive information. During the robustness test, we will assess our model on other data sizes. After hydrating the dataset using Twitter APIs, we obtain 570,464 user records and 920,281 tweets linked to 12,884 labeled claims. We preprocess the data by eliminating NaN values, removing duplicates, and filtering out non-English tweets, yielding a cleaned dataset comprising 388,561 tweets and 570,464 users. We then merge the tweets and fact-checked claims to derive 12,300 labeled tweets, among which 11,464 are categorized as misinformation.

We construct the User-Tweet network by treating users and tweets as nodes. In terms of edges, User—User edges are established based on following and mentioning relationships, while Tweet—Tweet edges are formed if they originate from the same claim or share cluster keywords. User—Tweet edges are created if a user posts, retweets, replies to, or quotes a tweet. We then employ the aforementioned fine-tuned stance detection model to label the stances of available edges.

Our goal is to predict if a user will propagate misinformation in the subsequent time period. If a link between the user and misinformation emerges in the next period, we label it as 1, otherwise 0. Since the graph evolves over time, we strictly adhere to the timestamp when dividing the data into train/validation/test sets (8:1:1) to prevent data leakage issues. As depicted in Figure 3, we employ the graph structure from an earlier time period (i.e., Time $t$) to train the model and subsequently predict potential propagation behaviors (i.e., the dashed line) in the following time period (i.e., Time $t + L$).

Our model leverages both node-level and network-level features to forecast misinformation dissemination. Thus, we select benchmarks based on feature type usage. For nodal features, we use two word embedding techniques to convert textual features (e.g., tweet content, user description) into vectors. The first technique is the widely adopted pre-trained word2vec model, GloVe, while the second is RoBERTa, a large language model known for its strong transfer learning capabilities (Liu et al., 2019; Pennington et al., 2014). For network-level features, we apply the node2vec model, which uses random walks to embed network nodes into vectors, ensuring that closely connected nodes in the network are also closed in vector space (Grover & Leskovec, 2016).

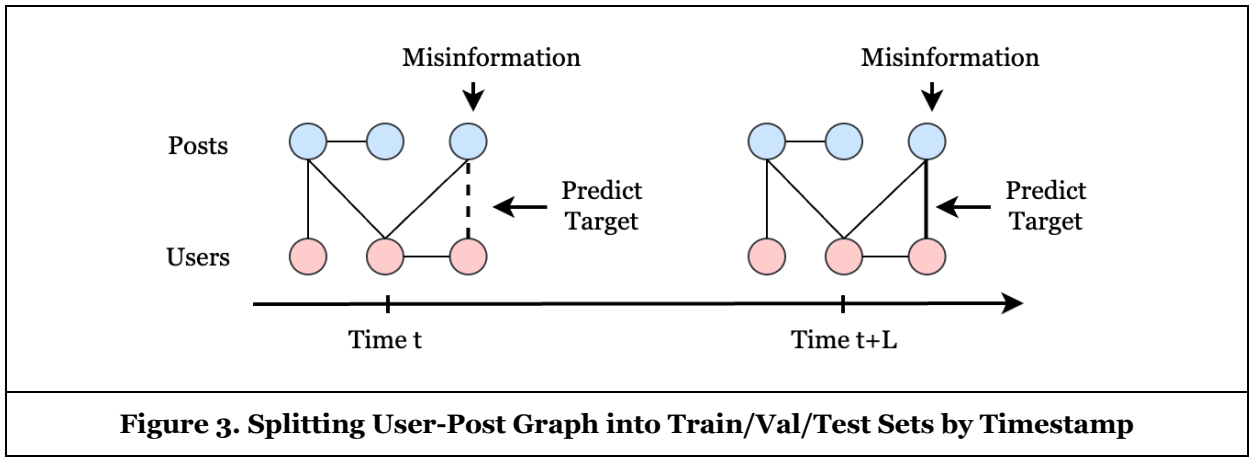

**Figure 3. Splitting User-Post Graph into Train/Val/Test Sets by Timestamp**

While node2vec excels at handling homogeneous graphs, it is less effective for heterogeneous graphs with multiple types of nodes and edges. To address this gap, we employ metapath2vec, an algorithm specifically designed for heterogeneous networks (Dong et al., 2017). Unlike node2vec's generic random walks, metapath2vec uses a predefined metapath schema to guide its random walks, enabling a more refined characterization of relationships among various node types. In our study on user-post interactions, we formulate metapaths such as ("user", "follows", "user") and ("user", "posts", "tweet") to guide the direction of random walks and better capture the relations between users and tweets.

To combine features derived from both nodal attributes and network structures, we integrate RoBERTa with node2vec and metapath2vec: Initially, we collect feature vectors generated by RoBERTa and node2vec/metapath2vec. Given that these feature sets inherently reside in distinct vector spaces, we pass the features through respective neural networks for linear transformation, which projects these high-dimensional vectors into a common latent space. Subsequently, the transformed vectors are concatenated and processed through an additional neural network layer, yielding the integrated feature representations. With these unified embeddings in hand, we select three distinct classifiers for final prediction tasks: Logistic Regression, Support Vector Classification, and Neural Networks.

Lastly, we include four GNN models without stance-aware paths in the baseline: Heterograph-GCN, Heterograph-GraphSAGE, Heterograph-GAT, and Transformer-based GNN. All these benchmark GNNs are adapted for heterogeneous graphs, and we have ensured they possess comparable parameter sizes. Notably, since both Heterograph-GAT and Transformer-based GNN leverage attention mechanisms to bolster model performance, the comparison between stance-aware GNN and these two models underscores the efficacy of the stance-informed metapath in predicting user behaviors upon encountering misinformation. Following previous IS literature, we employ AUC as our primary metric to evaluate the model's performance on out-of-sample data (Ma et al. 2021).

## Results

This section presents the results of our experiment aimed at predicting whether a user will spread misinformation on Twitter. The table below offers a comprehensive comparison of model performances on the test set in terms of AUC. The columns Text and Network indicate which features the model used in the experiment, while the Improvement column demonstrates the extent to which our proposed model surpasses each baseline. Furthermore, since we select three classifiers for each embedding, we only report the best-performing classifier in the AUC Score column of the table.

| Model Type | Text | Network | AUC Score | Improvement |
|---|---|---|---|---|
| Word2Vec | ✓ | | 0.5063 | 81.99% |
| RoBERTa | ✓ | | 0.7175 | 28.42% |
| Node2Vec | | ✓ | 0.5024 | 83.40% |
| Metapath2Vec | | ✓ | 0.5133 | 79.51% |
| Node2Vec + RoBERTa | ✓ | ✓ | 0.6722 | 37.07% |
| Metapath2Vec + RoBERTa | ✓ | ✓ | 0.6946 | 32.65% |
| Heterograph-GCN | ✓ | ✓ | 0.8650 | 6.52% |
| Heterograph-GraphSAGE | ✓ | ✓ | 0.8713 | 5.75% |
| Heterograph-GAT | ✓ | ✓ | 0.8775 | 5.01% |
| Transformer-based GNN | ✓ | ✓ | 0.8801 | 4.69% |
| GNN (with Stance-Aware Paths) | ✓ | ✓ | 0.9214 | —— |

**Table 1. The AUC Comparison between Stance-Aware GNN and Benchmarks**

First, among the models relying solely on node features, RoBERTa achieves the best performance with an AUC score of 0.7175, while Word2Vec lags behind with 0.5063. A key reason for this disparity is the prevalence of typos and word abbreviations in tweets, which are effectively addressed by deep learning models trained on tokens rather than complete words (Liu et al., 2019). In contrast, pre-trained word2vec models struggle in such scenarios. Moreover, when examining models that solely employ network features, both Node2Vec and Metapath2Vec yield relatively low AUC scores, indicating a difficulty in differentiating between users who spread misinformation and those who do not. Although Metapath2vec slightly outperforms node2vec due to its ability to incorporate additional information, both models fail to capture the requisite structural information for misinformation prediction.

Second, our results demonstrate that the integration of both node and network features leads to enhanced performance compared to using a single feature. Specifically, both Node2Vec + RoBERTa and Metapath2Vec + RoBERTa models achieve higher AUC scores (0.6722 and 0.6946, respectively) compared to their single-feature counterparts. These results align with the literature suggesting that combining different types of features can lead to more accurate predictions (Baltrušaitis et al., 2019). Utilizing the same two-feature combination, GAT and GraphSAGE demonstrate strong performance among other benchmarks in predicting misinformation dissemination, emphasizing the potency of GNNs.

Third, our stance-aware GNN significantly outperforms benchmark models in predicting misinformation spread. Compared to the four GNNs adapted for heterogeneous graphs but lacking echo-chamber driven paths, stance-aware GNN exhibits a performance boost of at least 4.69%. This finding aligns with Friggeri et al. (2014), who highlight the merits of integrating social influence into models for more accurate predictions of information spread. It is noteworthy that, while Heterograph-GAT and Transformer-based GNN encompass the same attention mechanisms, node features, and adjacent matrices, stance-aware GNN still emerges superior courtesy of its stance-based paths. This superiority resonates with previous research suggesting that the original adjacency relations between nodes may not always be optimal and can constrain prediction efficacy (Gong et al., 2019 CVPR). Thus, these stance-based paths act as a filter during the node attribute aggregation process, effectively sifting out noise from less relevant neighbors. In other words, stance-based metapaths enable users to selectively engage with neighbors who are influential in shaping their reactions to misinformation, which in turn enhances the predictive accuracy of models. This notion is further substantiated by the work of Bakshy et al. (2012), which shows that users are more likely to be influenced by information from close connections. Our study is the first to incorporate users' stances and social influence into deep learning models, enabling the model to better capture the intricate dynamics of social influence and information diffusion within online communities.

The above prediction results demonstrate the usefulness of user stances and the resulting network structures in the context of misinformation spread. To gain a deeper understanding, we further investigate two questions: are all types of structures equally important? Which type of network structure is the most influential one? The attention scores in our model provide the explainability, which allows us to examine the underlying mechanisms. As shown in the graph below, we initialize the trainable attention weights of four information propagation paths equally. Throughout the training process, the model adjusts the parameters to optimize predictive performance, and the line in the graph illustrates the weight changes for each path type, which represents the corresponding echo chamber structure.

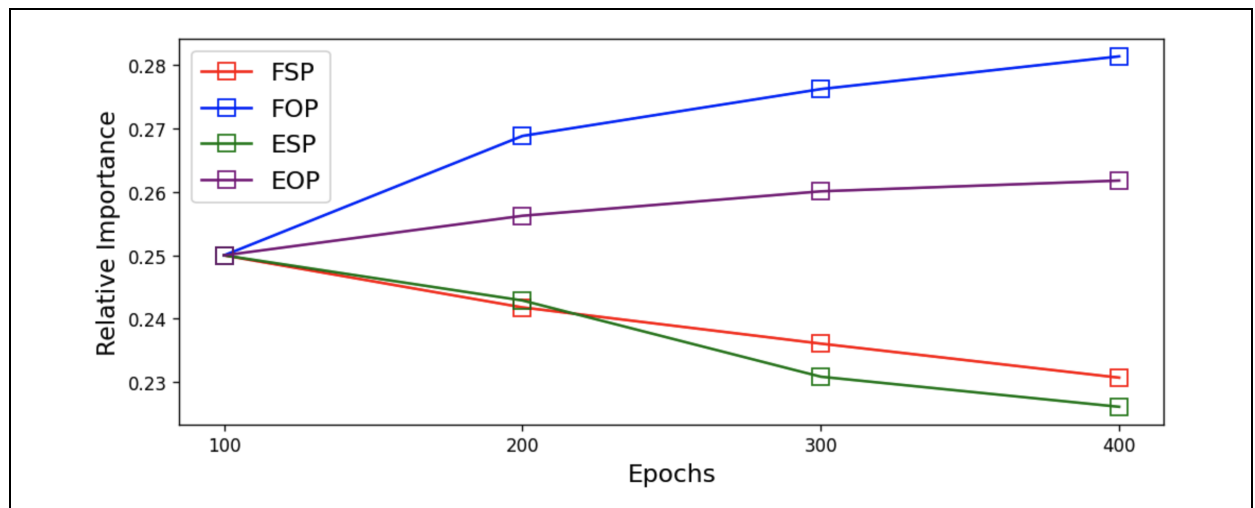

**Figure 4. Importance Evolution of Information Passing Paths during Training**

We observe two key findings in Figure 4. First, the weights of FOP and EOP increase over the training, while those of FSP and ESP decline. This indicates that an opposing stance offers a stronger signal than a supportive one. In other words, when a user demonstrates an opposing stance towards misinformation, it serves as a potent indicator that their neighbors are less likely to propagate the false information. Consider a user who follows two other users, both of whom repost a piece of misinformation but with differing stances—one supports it, while the other opposes it. Our results imply that the user in question is more likely to hinder the spread of misinformation, as the opposing stance carries greater significance. Second, we find that follower-based network structure has a more significant impact on the model compared to engagement-based structure. This finding aligns with the network homophily theory, which posits that users with similar characteristics may follow one another and form strong network ties (McPherson et al. 2001). In such a scenario, one user's influence can propagate more easily, swaying the stances of other users and, consequently, impacting the spread of misinformation.

Finally, as the original dataset offers three data sizes (small, medium, large), we perform a robustness test by deploying our model on networks that are smaller and have sparser connections. As expected, model performance declines as data size diminishes, primarily because smaller datasets provide less information for training, and deep learning models are typically data-hungry. Nevertheless, our model consistently outperforms the best benchmarks across all scenarios, with the performance gap widening as the data size increases. This demonstrates the consistency and capacity of our model, indicating its potential for robust performance when applied to more extensive data streams in real-world settings.

In summary, our experimental results indicate that our stance-aware GNN significantly outperforms all the benchmarks in predicting misinformation spread on Twitter. The integration of user stance and social influence in our model allows for a more accurate representation of the complex dynamics of information diffusion in online networks. Moreover, we find that opposing stances and follower-based social network structure serves as a stronger signal in misinformation prediction. Lastly, the robustness test further highlights the consistency and capacity of our model, suggesting its potential for handling larger data streams in real-world applications.

## Discussion

In this study, we introduce the stance-aware GNN, a novel approach that integrates user stances and network mechanisms to predict misinformation spread on social media platforms. Recognizing the potential value of ordinary users' opinions in misinformation spread, we designed four types of information propagation paths that integrated echo chambers structures and different user stances. We also implemented trainable weights that demonstrated the relative significance of each path, thereby reflecting the different influences of echo chamber structures. Our experiments with a real-world dataset reveal that the stance-aware GNN outperforms traditional benchmark models and GNNs without user stance. The attention weights further demonstrated that echo chamber structures with opposing stances are more influential in explaining misinformation spread. Our findings show that ordinary users' stances can be social signals that influence misinformation spread. Our artifact thereby offers a valuable predictive instrument for platforms and organizations aiming to tackle misinformation dissemination.

Our research contributes to the ongoing conversation of misinformation spread mitigation in two primary ways. First, we demonstrate the vital role of user stances in the formation of echo chambers and their subsequent impact on misinformation diffusion. As existing research on top-down platform intervention shows mixed findings, it is important for researchers to consider how to mobilize ordinary social media users as crowdsourced social debunking efforts against misinformation (Borwankar et al., 2022; Vraga & Bode, 2017). Extending current research on the effectiveness of social media users' opinions, our findings corroborate how user stance may affect the behaviors of local neighborhoods. Although the echo chamber is a well-received concept of information consumption in social media, we unravel more nuanced mechanisms by integrating both network content and network structure, which is especially important for social media networks (Kane et al., 2014). By examining the attention weights among different paths, our findings reveal the critical role of opposing user stances in influencing echo chamber mechanisms, which illustrates the values of social correction.

Furthermore, our work follows the network paradigm that views information propagation as a social process with inherent dependency mechanisms. These dependency mechanisms are local organizing processes that give rise to a global structure, that is, the global structure emerges through bottom-up patterns involving only local interactions. In the context of misinformation spread, these local interactions describe how users respond to others' interpretations and form/dissemble clusters. We represent such local interactions as paths in GNN, and the significant model improvement indicates that the stance-integrated echo chamber mechanisms are prominent social mechanisms in misinformation spread. Hence, the stance-aware GNN model not only has higher predictive power but better explainability.

As one of the pioneering studies in the IS field that incorporates user stance and graph neural networks to predict misinformation spread in social networks, the paper is not without its limitations. First, our investigation primarily focuses on users' self-descriptions and post content as features. While these are salient features, past research underscores the supplementary information, such as user identity verification, can be valuable for discerning users' propensity for disseminating fake news (S. Wang et al., 2021). The heterogeneity in user demographics and posting behavior across diverse platforms (Yang et al., 2019) also suggests that the integration of user stance distributions might bolster the efficacy of stance-aware GNN. Another important additional feature is the misinformation tagging provided by online platforms. Although prior IS research presents a mixed view on the utility of such platform-driven tags (Moravec et al., 2019; Ng et al., 2021), exploring their synergy with user-level social corrections promises to be a fertile area of inquiry.

Second, the time-relevant information merits deeper exploration. In our current model, we split the dataset following the timestamp of users' posts and predict if a user will circulate misinformation in a subsequent time period. Future studies could probe how variations in the prediction horizon—extending or reducing the time frame—affect the model's efficacy. From a model architecture standpoint, our current design encapsulates four message passing paths that resonate with distinct echo chamber structures throughout the entire training phase. Future research might study more dynamic structures such as charting the evolution of echo chamber structures. Moreover, as previous studies have illuminated the waning influence of users on their neighbors over time, we can also assign time-decay weights to edges based on temporal proximity. Further granularity in prediction, like forecasting a user's evolving stance towards misinformation, also offers a compelling avenue for exploration.

The final limitation relates to our stance detection model, which, despite being trained on a comprehensive sentence-pair dataset with stance labels, achieves an accuracy of only 82.3%. This limitation could potentially introduce biases to the model results, given that certain stance-based paths might be mislabeled. To mitigate this, we suggest employing larger language models (e.g., GPT-4 or Llamma2) and exploring prompt engineering to understand stance determination rationale. These steps not only enhance model transparency but also improve its accountability (Bauer et al., 2023; Berente et al., 2021).

## Conclusion

Our research illuminates the integrative role of user stances and network mechanisms in misinformation spread, providing an effective prediction tool for platforms, organizations, and policymakers seeking to address the challenges posed by misinformation dissemination. While the stance-aware GNN represents a significant step towards utilizing the power of deep learning and network analysis in the ongoing battle against misinformation, there are still numerous promising avenues for future research to expand upon this work. Addressing these limitations and exploring the proposed directions hold the potential to further contribute to the IS field in both theoretical and practical dimensions, enhancing our understanding of misinformation spread and fostering the development of increasingly sophisticated countermeasures.

## References

Bakshy, E., Rosenn, I., Marlow, C., & Adamic, L. (2012). The role of social networks in information diffusion. Proceedings of the 21st International Conference on World Wide Web, 519–528.

Baltrušaitis, T., Ahuja, C., & Morency, L.-P. (2019). Multimodal Machine Learning: A Survey and Taxonomy. IEEE Transactions on Pattern Analysis and Machine Intelligence, 41(2), 423–443.

Bauer, K., von Zahn, M., & Hinz, O. (2023). Expl(AI)ned: The Impact of Explainable Artificial Intelligence on Users' Information Processing. Information Systems Research.

Benbya, H., Davenport, T. H., & Pachidi, S. (2020). Artificial intelligence in organizations: Current state and future opportunities. MIS Quarterly Executive, 19(4).

Berente, N., Gu, B., Recker, J., & Santhanam, R. (2021). Special Issue Editor's Comments: Managing Artificial Intelligence. Management Information Systems Quarterly, 45(3), 1433–1450.

Borwankar, S., Zheng, J., & Kannan, K. N. (2022). Democratization of Misinformation Monitoring: The Impact of Twitter's Birdwatch Program (SSRN Scholarly Paper No. 4236756).

Clarke, J., Chen, H., Du, D., & Hu, Y. J. (2021). Fake News, Investor Attention, and Market Reaction. Information Systems Research, 32(1), 35–52.

Conforti, C., Berndt, J., Pilehvar, M. T., Giannitsarou, C., Toxvaerd, F., & Collier, N. (2020). Will-They-Won't-They: A Very Large Dataset for Stance Detection on Twitter. Proceedings of the 58th Annual Meeting of the Association for Computational Linguistics, 1715–1724.

Deng, B., & Chau, M. (2021). The Effect of the Expressed Anger and Sadness on Online News Believability. Journal of Management Information Systems, 38(4), 959–988.

Dong, Y., Chawla, N. V., & Swami, A. (2017). metapath2vec: Scalable Representation Learning for Heterogeneous Networks. Proceedings of the 23rd ACM SIGKDD International Conference on Knowledge Discovery and Data Mining, 135–144.

Festinger, L. (1962). Cognitive Dissonance. Scientific American, 207(4), 93–106.

Friggeri, A., Adamic, L., Eckles, D., & Cheng, J. (2014). Rumor Cascades. Proceedings of the International AAAI Conference on Web and Social Media, 8(1), Article 1.

George, J., Gerhart, N., & Torres, R. (2021). Uncovering the Truth about Fake News: A Research Model Grounded in Multi-Disciplinary Literature. Journal of Management Information Systems, 38(4), 1067–1094.

Gorrell, G., Bontcheva, K., Derczynski, L., Kochkina, E., Liakata, M., & Zubiaga, A. (2018). RumourEval 2019: Determining Rumour Veracity and Support for Rumours (arXiv:1809.06683). arXiv.

Grover, A., & Leskovec, J. (2016). node2vec: Scalable Feature Learning for Networks. ArXiv:1607.00653 [Cs, Stat].

Habernal, I., Wachsmuth, H., Gurevych, I., & Stein, B. (2018). The Argument Reasoning Comprehension Task: Identification and Reconstruction of Implicit Warrants. Proceedings of the 2018 Conference of the North American Chapter of the Association for Computational Linguistics: Human Language Technologies, Volume 1 (Long Papers), 1930–1940.

Hamilton, W., Ying, Z., & Leskovec, J. (2017). Inductive Representation Learning on Large Graphs. Advances in Neural Information Processing Systems, 30.

Heider, F. (1982). The Psychology of Interpersonal Relations. Psychology Press.

Horner, C. G., Galletta, D., Crawford, J., & Shirsat, A. (2021). Emotions: The Unexplored Fuel of Fake News on Social Media. Journal of Management Information Systems, 38(4), 1039–1066.

Jasny, L., Dewey, A. M., Robertson, A. G., Yagatich, W., Dubin, A. H., Waggle, J. M., & Fisher, D. R. (2018). Shifting echo chambers in US climate policy networks. PLOS ONE, 13(9), e0203463.

Kane, G. C., Alavi, M., Labianca, G. (Joe), & Borgatti, S. P. (2014). What'S Different About Social Media Networks? A Framework and Research Agenda. MIS Quarterly, 38(1), 275–304.

Kim, A., & Dennis, A. R. (2019). Says who? The effects of presentation format and source rating on fake news in social media. MIS Quarterly, 43(3), 1025–1039.

Kim, A., Moravec, P. L., & Dennis, A. R. (2019). Combating Fake News on Social Media with Source Ratings: The Effects of User and Expert Reputation Ratings. Journal of Management Information Systems, 36(3), 931–968.

King, K. K., Wang, B., Escobari, D., & Oraby, T. (2021). Dynamic Effects of Falsehoods and Corrections on Social Media: A Theoretical Modeling and Empirical Evidence. Journal of Management Information Systems, 38(4), 989–1010.

Kipf, T. N., & Welling, M. (2017). Semi-Supervised Classification with Graph Convolutional Networks. ArXiv:1609.02907 [Cs, Stat].

Kitchens, B., Johnson, S. L., & Gray, P. (2020). Understanding Echo Chambers and Filter Bubbles: The Impact of Social Media on Diversification and Partisan Shifts in News Consumption. MIS Quarterly, 44(4), 1619–1649.

Lazer, D. M., Baum, M. A., Benkler, Y., Berinsky, A. J., Greenhill, K. M., Menczer, F., Metzger, M. J., Nyhan, B., Pennycook, G., & Rothschild, D. (2018). The science of fake news. Science, 359(6380), 1094–1096.

LeCun, Y., Bengio, Y., & Hinton, G. (2015). Deep learning. Nature, 521(7553), Article 7553.

Liu, Y., Ott, M., Goyal, N., Du, J., Joshi, M., Chen, D., Levy, O., Lewis, M., Zettlemoyer, L., & Stoyanov, V. (2019). RoBERTa: A Robustly Optimized BERT Pretraining Approach (arXiv:1907.11692). arXiv.

MA, J., GAO, W., MITRA, P., KWON, S., JANSEN, B. J., WONG, K.-F., & CHA, M. (2016). Detecting rumors from microblogs with recurrent neural networks. Proceedings of the 25th International Joint Conference on Artificial Intelligence (IJCAI 2016), 3818–3824.

Ma, T., Hu, Y., Lu, Y., & Bhattacharyya, S. (2021). Graph Neural Network for Customer Engagement Prediction on Social Media Platforms. Hawaii International Conference on System Sciences 2021 (HICSS-54).

McKimmie, B. M., Terry, D. J., Hogg, M. A., Manstead, A. S. R., Spears, R., & Doosje, B. (2003). I'm a hypocrite, but so is everyone else: Group support and the reduction of cognitive dissonance. Group Dynamics: Theory, Research, and Practice, 7, 214–224.

McPherson, M., Smith-Lovin, L., & Cook, J. M. (2001). Birds of a Feather: Homophily in Social Networks. Annual Review of Sociology, 27(1), 415–444.

Mohammad, S., Kiritchenko, S., Sobhani, P., Zhu, X., & Cherry, C. (2016). SemEval-2016 Task 6: Detecting Stance in Tweets. Proceedings of the 10th International Workshop on Semantic Evaluation (SemEval-2016), 31–41.

Monge, P. R., & Contractor, N. (2003). Theories of Communication Networks. Oxford University Press.

Moravec, P. L., Kim, A., & Dennis, A. R. (2020). Appealing to Sense and Sensibility: System 1 and System 2 Interventions for Fake News on Social Media. Information Systems Research, 31(3), 987–1006.

Moravec, P. L., Kim, A., Dennis, A. R., & Minas, R. K. (2022). Do You Really Know if It's True? How Asking Users to Rate Stories Affects Belief in Fake News on Social Media. Information Systems Research, 33(3), 887–907.

Moravec, P. L., Minas, R. K., & Dennis, A. (2019). Fake News on Social Media: People Believe What They Want to Believe When it Makes No Sense At All. MIS Quarterly, 43(4), 1343–1360.

Ng, K. C., Tang, J., & Lee, D. (2021). The Effect of Platform Intervention Policies on Fake News Dissemination and Survival: An Empirical Examination. Journal of Management Information Systems, 38(4), 898–930.

Nickerson, R. S. (1998). Confirmation Bias: A Ubiquitous Phenomenon in Many Guises. Review of General Psychology, 2(2), 175–220.

Nielsen, D. S., & McConville, R. (2022). MuMiN: A Large-Scale Multilingual Multimodal Fact-Checked Misinformation Social Network Dataset. Proceedings of the 45th International ACM SIGIR Conference on Research and Development in Information Retrieval, 3141–3153.

O'Connor, C., & Weatherall, J. O. (2019). The Misinformation Age: How False Beliefs Spread. Yale University Press.

Oh, O., Agrawal, M., & Rao, H. R. (2013). Community Intelligence and Social Media Services: A Rumor Theoretic Analysis of Tweets During Social Crises. MIS Quarterly, 37(2), 407–426.

Pennington, J., Socher, R., & Manning, C. (2014). GloVe: Global Vectors for Word Representation. Proceedings of the 2014 Conference on Empirical Methods in Natural Language Processing (EMNLP), 1532–1543.

Pennycook, G., Bear, A., Collins, E. T., & Rand, D. G. (2020). The Implied Truth Effect: Attaching Warnings to a Subset of Fake News Headlines Increases Perceived Accuracy of Headlines Without Warnings. Management Science, 66(11), 4944–4957.

Rosenthal, S., Farra, N., & Nakov, P. (2019). SemEval-2017 Task 4: Sentiment Analysis in Twitter (arXiv:1912.00741). arXiv.

Ross, B., Jung, A., Heisel, J., & Stieglitz, S. (2018). Fake News on Social Media: The (In)Effectiveness of Warning Messages. Proceedings of the 39th International Conference on Information Systems (ICIS 2018), 16.

Saeed, M., Traub, N., Nicolas, M., Demartini, G., & Papotti, P. (2022). Crowdsourced Fact-Checking at Twitter: How Does the Crowd Compare With Experts? Proceedings of the 31st ACM International Conference on Information & Knowledge Management, 1736–1746.

Shore, J., Baek, J., & Dellarocas, C. (2018). Network Structure and Patterns of Information Diversity on Twitter. Management Information Systems Quarterly, 42(3), 849–972.

Song, H., Cho, J., & Benefield, G. A. (2020). The Dynamics of Message Selection in Online Political Discussion Forums: Self-Segregation or Diverse Exposure? Communication Research, 47(1), 125–152.

Stroebe, W., & Diehl, M. (1981). Conformity and counterattitudinal behavior: The effect of social support on attitude change. Journal of Personality and Social Psychology, 41, 876–889.

Turel, O., & Osatuyi, B. (2021). Biased Credibility and Sharing of Fake News on Social Media: Considering Peer Context and Self-Objectivity State. Journal of Management Information Systems, 38(4), 931–958.

Vosoughi, S., Roy, D., & Aral, S. (2018). The spread of true and false news online. Science, 359(6380), 1146–1151.

Vraga, E. K., & Bode, L. (2017). Using Expert Sources to Correct Health Misinformation in Social Media. Science Communication, 39(5), 621–645.

Wang, S., Pang, M.-S., & Pavlou, P. A. (2021). Cure or Poison? Identity Verification and the Posting of Fake News on Social Media. Journal of Management Information Systems, 38(4), 1011–1038.

Wang, X., Ji, H., Shi, C., Wang, B., Cui, P., Yu, P., & Ye, Y. (2021). Heterogeneous Graph Attention Network (arXiv:1903.07293). arXiv.

Wu, Z., Pan, S., Chen, F., Long, G., Zhang, C., & Yu, P. S. (2021). A Comprehensive Survey on Graph Neural Networks. IEEE Transactions on Neural Networks and Learning Systems, 32(1), 4–24.

Xu, Y., Sun, Y., Hagen, L., Patel, M., & Falling, M. (2021). Evolution of the plandemic communication network among serial participants on Twitter. New Media & Society, 14614448211050928.

Yang, M., Ren, Y., & Adomavicius, G. (2019). Understanding User-Generated Content and Customer Engagement on Facebook Business Pages. Information Systems Research, 30(3), 839–855.

Yoo, E., Gu, B., & Rabinovich, E. (2019). Diffusion on Social Media Platforms: A Point Process Model for Interaction among Similar Content. Journal of Management Information Systems, 36(4), 1105–1141.

Yuan, H., Zheng, J., Ye, Q., Qian, Y., & Zhang, Y. (2021). Improving fake news detection with domain-adversarial and graph-attention neural network. Decision Support Systems, 151, 113633.